\documentclass{article}

\usepackage{microtype}
\usepackage{graphicx}
\usepackage{subfigure}
\usepackage{booktabs} 

\usepackage{hyperref}

\usepackage[accepted]{icml2024}

\usepackage{amsmath}
\usepackage{amssymb}
\usepackage{mathtools}
\usepackage{amsthm}

\usepackage[capitalize,noabbrev]{cleveref}

\theoremstyle{plain}

\theoremstyle{definition}

\theoremstyle{remark}

\usepackage[textsize=tiny]{todonotes}

\icmltitlerunning{ProtIR: Iterative Refinement between Retrievers and Predictors for Protein Function Annotation}

\usepackage{amsmath,amsfonts,bm}

\def\eqref#1{equation~\ref{#1}}

\def\1{\bm{1}}

\def\vc{{\bm{c}}}

\def\vh{{\bm{h}}}

\def\vv{{\bm{v}}}

\def\vx{{\bm{x}}}
\def\vy{{\bm{y}}}

\DeclareMathAlphabet{\mathsfit}{\encodingdefault}{\sfdefault}{m}{sl}
\SetMathAlphabet{\mathsfit}{bold}{\encodingdefault}{\sfdefault}{bx}{n}

\def\gK{{\mathcal{K}}}

\def\gN{{\mathcal{N}}}

\newcommand{\E}{\mathbb{E}}

\usepackage{enumitem}
\usepackage{xspace}
\usepackage{xcolor}
\usepackage{wrapfig}
\usepackage{caption}
\usepackage{amssymb}
\usepackage{multirow}
\usepackage{adjustbox}
\usepackage{subfigure}
\usepackage{makecell}
\usepackage{mathtools}
\usepackage{algorithm}
\usepackage{colortbl}
\usepackage{tcolorbox}
\usepackage{babel}
\usepackage{bbding}
\usepackage{pifont}
\usepackage{nicematrix}
\usepackage{enumitem}
\usepackage{booktabs}
\usepackage[flushleft]{threeparttable}

\usepackage{thmtools}
\usepackage{thm-restate}

\definecolor{myblue}{RGB}{68, 114, 196}
\definecolor{myorange}{RGB}{237, 125, 49}
\definecolor{lightgray}{RGB}{229, 232, 232}
\definecolor{mydarkblue}{rgb}{0,0.08,0.45}

\newcommand{\method}{ProtIR\xspace}

\usepackage{hyperref}
\usepackage{url}

\begin{document}

\twocolumn[
\icmltitle{ProtIR: Iterative Refinement between Retrievers and Predictors \\ for Protein Function Annotation}



\icmlsetsymbol{equal}{*}

\begin{icmlauthorlist}
\icmlauthor{Zuobai Zhang}{mila,udem}
\icmlauthor{Jiarui Lu}{mila,udem}
\icmlauthor{Vijil Chenthamarakshan}{ibm}
\icmlauthor{Aur\'{e}lie Lozano}{ibm}
\icmlauthor{Payel Das}{ibm}
\icmlauthor{Jian Tang}{mila,hec,cifar}
\end{icmlauthorlist}

\icmlaffiliation{mila}{Mila - Qu\'ebec AI Institute}
\icmlaffiliation{udem}{Universit\'e de Montr\'eal}
\icmlaffiliation{ibm}{IBM Research}
\icmlaffiliation{hec}{HEC Montr\'eal}
\icmlaffiliation{cifar}{CIFAR AI Chair}

\icmlcorrespondingauthor{Zuobai Zhang}{zuobai.zhang@mila.quebec}
\icmlcorrespondingauthor{Payel Das}{daspa@us.ibm.com}
\icmlcorrespondingauthor{Jian Tang}{jian.tang@hec.ca}

\icmlkeywords{Machine Learning, ICML}

\vskip 0.3in
]



\printAffiliationsAndNotice{}  


\begin{abstract}
Protein function annotation is an important yet challenging task in biology.
Recent deep learning advancements show significant potential for accurate function prediction by learning from protein sequences and structures.
Nevertheless, these predictor-based methods often overlook the modeling of protein similarity, an idea commonly employed in traditional approaches using sequence or structure retrieval tools.
To fill this gap, we first study the effect of inter-protein similarity modeling by benchmarking retriever-based methods against predictors on protein function annotation tasks. 
Our results show that retrievers can match or outperform predictors without large-scale pre-training.
Building on these insights, we introduce a novel variational pseudo-likelihood framework, \textbf{\method}, designed to improve function predictors by incorporating inter-protein similarity modeling. 
This framework iteratively refines knowledge between a function predictor and retriever, thereby combining the strengths of both predictors and retrievers.
\method showcases around 10\% improvement over vanilla predictor-based methods. 
Besides, it achieves performance on par with protein language model-based methods, yet without the need for massive pre-training, highlighting the efficacy of our framework.
Code will be released upon acceptance.
\end{abstract}
\section{Introduction}

\begin{figure}[t]
    \centering
    \includegraphics[width=\linewidth]{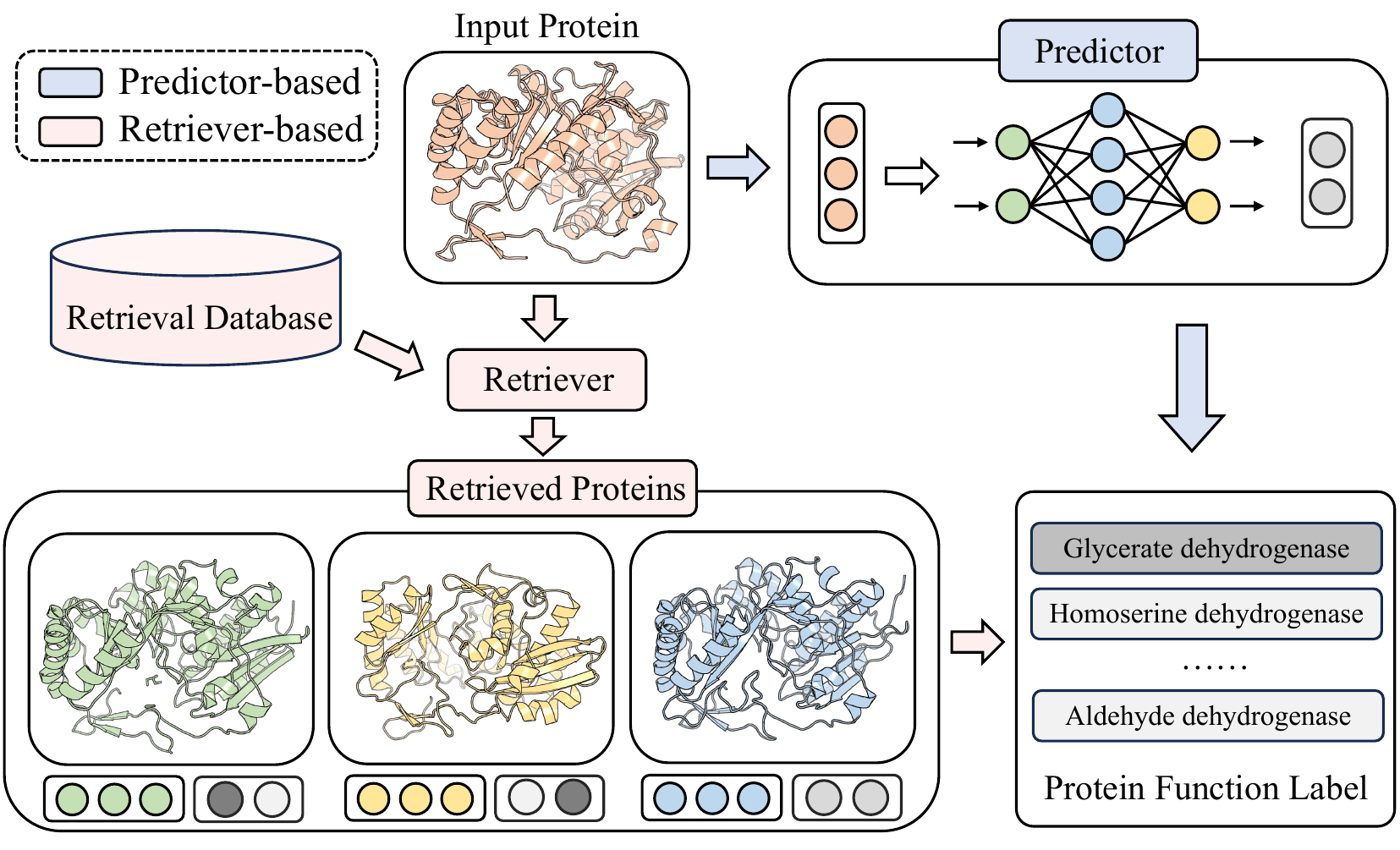}
    \caption{High-level illustration of predictor- and retriever-based methods for protein function annotation.}
    \label{fig:retriever}
\end{figure}

Proteins, being fundamental components in biological systems, hold a central position in a myriad of biological activities, spanning from catalytic reactions to cell signaling processes. 
The complexity of these macromolecules arises from the intricate interactions between their sequences, structures, and functionalities, influenced by both physical principles and evolutionary processes~\citep{Sadowski2009TheSR}. 
Despite decades of research, understanding protein function remains a challenge, with a large portion of proteins either lacking characterization or having incomplete understanding of their roles. 

Recent advancements in protein representation learning from sequences or structures have shown promise for accurate function prediction.
Among these approaches, sequence-based methods treat protein sequences as the language of life and train protein language models on billions of natural protein sequences~\citep{elnaggar2020prottrans,elnaggar2023ankh,rives2021biological,lin2023evolutionary}, while structure-based methods model protein structures as graphs and employ 3D graph neural networks to facilitate message passing between various residues~\citep{gligorijevic2021structure,zhang2022protein,fan2023continuousdiscrete}.

Despite the impressive performance of these machine learning techniques in predicting protein functions, these methods often neglect the modeling of protein similarity—a key idea used by traditional approaches. 
This modeling is achieved through the use of widely adopted sequence comparison tools such as BLAST~\citep{mcginnis2004blast,conesa2005blast2go}.
These tools operate under the evolutionary assumption that proteins with similar sequences likely possess similar functions, offering interpretability by identifying the most closely related reference example for function annotation~\citep{dickson2023fine}.
Beyond function annotation by retrieving similar sequences, another plausible assumption is that proteins with similar structures also exhibit similar functions, as protein structures directly determine function~\citep{Roy2015APO}.
Recent advances in structure retrievers~\citep{van2023fast}, along with progress in structure prediction protocols~\citep{jumper2021highly,lin2023evolutionary}, have paved the way to explore function annotation methods based on structure retrievers.
However, there is still a lack of studies investigating the performance of the modern protein retrievers compared to predictor-based methods for function annotation.

In this paper, we study the effect of inter-protein similarity modeling and introduce a flexible framework, \textbf{\method}, as a means to improve function predictors.
We first \emph{benchmark} various retriever-based methods against predictor-based approaches on standard protein function annotation tasks (Fig.~\ref{fig:retriever}), namely Enzyme Commission number and Gene Ontology term prediction.
Our evaluation includes both traditional and neural retrievers utilizing sequences and structures.
Experimental results show that retriever-based methods can yield comparable or superior performance compared to predictor-based neural approaches without massive pre-training.
However, without function-specific training, it remains a challenge for a retriever to match the state-of-the-art performance of predictor-based methods based on protein language models, regardless of whether the retriever is based on sequences or structures.


To exploit the effectiveness of retrievers, we next introduce an approach, named \method, that combines neural retrievers with neural predictors for function annotation.
We present an innovative variational pseudo-likelihood framework to model the joint distribution of functions over all labeled proteins, ultimately improving predictors without massive pre-training. 
Utilizing the EM algorithm to optimize the evidence lower bound, we develop an iterative refinement framework that iterates between function predictors and retrievers.
In the E-step, we keep the retriever fixed while training the predictor to mimic the labels inferred from the retriever.
Correspondingly, during the M-step, we freeze the predictor and fine-tune the retriever using the labels inferred from the predictor as the target. 
This process iteratively distills knowledge of the predictor and retriever into each other and can be applied to any protein encoder.
Our experimental results on two state-of-the-art protein structure encoders, GearNet~\citep{zhang2022protein} and CDConv~\citep{fan2023continuousdiscrete}, clearly demonstrate that the {\method} framework improves vanilla predictors by an average improvement of approximately 10\% across different datasets. Moreover, it achieves comparable performance to protein language model-based methods without pre-training on millions of sequences, underscoring the efficacy of our approach.
Our contributions are two-fold:
\begin{enumerate}[leftmargin=*,itemsep=2pt,topsep=0pt,parsep=0pt,partopsep=0pt]
    \item We systematically evaluate retriever- and predictor-based methods and highlight the effectiveness of retriever-based methods on protein function annotation.

    \item We formulate an iterative refinement framework  between predictors and retrievers, ProtIR, which significantly enhances predictors without massive pre-training.

\end{enumerate}

\section{Protein Function Annotation}

\subsection{Background}
\label{sec:def}

\textbf{Proteins.}
Proteins are macromolecules formed through the linkage of residues via peptide bonds.
While only 20 standard residue types exist, their exponential combinations play a pivotal role in the extensive diversity of proteins found in the natural world.
The specific ordering of these residues determines the 3D positions of all the atoms within the protein, \emph{i.e.}, the protein structure. 
Following the common practice, we utilize only the alpha carbon atoms to represent the backbone structure of each protein. 
Each protein $\vx$ can be seen as a pair of a sequence and structure,
and is associated with function labels $\vy \in \{0,1\}^{n_c}$, where there are $n_c$ distinct functional terms, and each element indicates whether the protein performs a specific function.

\textbf{Problem Definition.}
In this paper, we delve into the problem of protein function annotation. 
Given a set of proteins $\vx_V=\vx_L\bigcup \vx_U$ and the labels $\vy_L$ of a set of labeled proteins $L\subset V$, our objective is to predict the labels $\vy_U$ for the remaining unlabeled set $U= V \backslash L$.
Typically, methods based on supervised learning train an encoder denoted as $\psi$ to maximize the log likelihood $p_\psi(\vy_L|\vx_L)$ of the ground truth labels in the training set, known as \textbf{predictor-based methods}. This optimization can be formulated as:
\begin{equation}
\label{eq:predictor}
\begin{split}
    \max\nolimits_\psi\; &\begin{matrix}\sum\nolimits_{n\in L} \vy_{n} \log p_\psi(\vy_{n}|\vx_n)\end{matrix}\\
    +& \begin{matrix}\sum\nolimits_{n\in L} (\mathbf{1}-\vy_{n}) \log (\mathbf{1} - p_\psi(\vy_{n}|\vx_n))\end{matrix},
\end{split}
\end{equation}
where $p_\psi(\vy_n|\vx_n) = \sigma(\text{MLP}(\psi(\vx_n)))$, and $\sigma(\cdot)$ is the sigmoid function. 
The goal is to generalize the knowledge learned by the encoder to unlabeled proteins and maximize the likelihood $p_\psi(\vy_U|\vx_U)$ for unobserved function labels.


\subsection{Retriever-Based Function Annotation}
\label{sec:func_retriever}

Despite the success of predictor-based methods in protein function prediction, practical annotation often uses sequence alignment tools like BLAST~\citep{altschul1997gapped,conesa2005blast2go}. 
These methods infer sequence homology, \emph{i.e.}, common evolutionary ancestry from sequence similarity, which can further be connected with functional similarity~\citep{pearson2013introduction}, thus offering interpretability for function annotation.
These \textbf{retriever-based methods} exhibit a close connection with kernel methods commonly studied in machine learning~\citep{ShaweTaylor2003KernelMF}.
In this context, the prediction for an unlabeled protein $i\in U$ leverages the labels from the labeled set $L$ via the following expression:
\begin{equation}
\label{eq:retriver}
\begin{split}
&\hat{\vy}_i =\begin{matrix} {\sum_{j\in\gN_k(i)} \widetilde{\gK}(\vx_i,\vx_j) \cdot \vy_j},\end{matrix} \\
\;\text{with}\; &\begin{matrix}\widetilde{\gK}(\vx_i,\vx_j) = \gK(\vx_i,\vx_j)/ {\sum_{t\in\gN_k(i)} \gK(\vx_i,\vx_t)}\end{matrix}
\end{split}
\end{equation}
where the kernel function $\gK(\cdot,\cdot)$ quantifies the similarity between two proteins, and $\gN_k(i)\subset L$ represents the top-$k$ most similar proteins to protein $i$ in the labeled set. 
For efficiency, we consider only a subset of labeled proteins and re-normalize the similarity within the retrieved set $\gN_k(i)$. 
It is important to note that various methods differ in their specific definitions of the similarity kernel.

\subsection{Neural Structure Retriever}
\label{sec:struct_retriever}

Besides the sequence retrievers mentioned above, relations between structure and function similarity can be characterized by quantitative models, QSAR~\citep{Roy2015APO}, due to the direct impact of structures on functions. Recent developments in structure retrievers and prediction protocols like AlphaFold2~\citep{jumper2021highly} have opened up promising avenues for exploring structure-based retrieval methods, \emph{e.g.}, Foldseek~\citep{van2023fast}.


Moving beyond traditional retrievers that compare protein structures in Euclidean space, we now present a strategy to adopt advanced protein structure representation learning techniques. Our method adopts a protein structure encoder to map proteins into a high-dimensional latent space, where their similarities are measured using cosine similarity. To guarantee that these representations reflect structural information, we pre-train the encoder on a fold classification task~\citep{hou2018deepsf} using 16,712 proteins from 1,195 different folds in the SCOPe 1.75 database~\citep{murzin1995scop}. This pre-training helps ensure proteins within the same fold are  closely encoded in the latent space.


Formally, our goal is to learn a protein encoder $\phi$ through pre-training on a protein database $\vx_D$ with associated fold labels $\vc_D$. The training objective involves maximizing the conditional log likelihood:
\begin{equation}
\label{eq:fold}
\begin{split}
    &\max\nolimits_\phi\; \log p_\phi(\vc_D|\vx_D) \\
    = &\begin{matrix}\sum\nolimits_{n\in D} \sum_{c} [c_{n}=c] \log p_\phi(c_{n}=c|\vx_n)\end{matrix}.
\end{split}
\end{equation}
Subsequently, we define the kernel function in (\ref{eq:retriver}) as a Gaussian kernel on the cosine similarity:
\begin{equation}
\setlength{\abovedisplayskip}{3pt}
\setlength{\belowdisplayskip}{3pt}
\label{eq:kernel}
    \gK(\vx_i,\vx_j)=\exp(\text{cos}(\phi(\vx_i),\phi(\vh_j))/\tau),
\end{equation}
where $\tau$ serves as the temperature parameter, controlling the scale of similarity values and is typically set to $0.03$ in practice. In this work, we will consider GearNet~\citep{zhang2022protein} and CDConv~\citep{fan2023continuousdiscrete} as our choice of encoder $\phi$.
A notable advantage of neural retrievers over traditional ones is their flexibility in fine-tuning for specific functions, motivating the framework in the next section.

\section{ProtIR: Iterative Refinement Between Predictor and Retriever}
\label{sec:protir}
Retriever-based methods offer interpretable function prediction, but face challenges in accurately predicting all functions due to the diverse factors influencing protein functions. 
Predictor-based methods, on the other hand, excel by using labeled data to learn and predict functions for new proteins, yet confined to specific task and limited data. 
To combine the advantages of both, in this section, we introduce an iterative refinement framework based on the EM algorithm, alternating between function predictors and retrievers. 
In the E-step, we fix the retriever $\phi$ while allowing the predictor $\psi$ to mimic the labels inferred from the retriever, improving the precision of function annotation with inter-protein similarity.
In the M-step, we freeze the predictor $\psi$ and optimize the retriever $\phi$ with the labels inferred from the predictor as the target, effectively distilling the predictor's global protein function knowledge into the retriever.
This collaborative process mutually strengthens the performance of both the predictor and retriever.



\subsection{A Pseudolikelihood Variational EM Framework}

\begin{figure*}[t]
    \centering
    \includegraphics[width=0.8\linewidth]{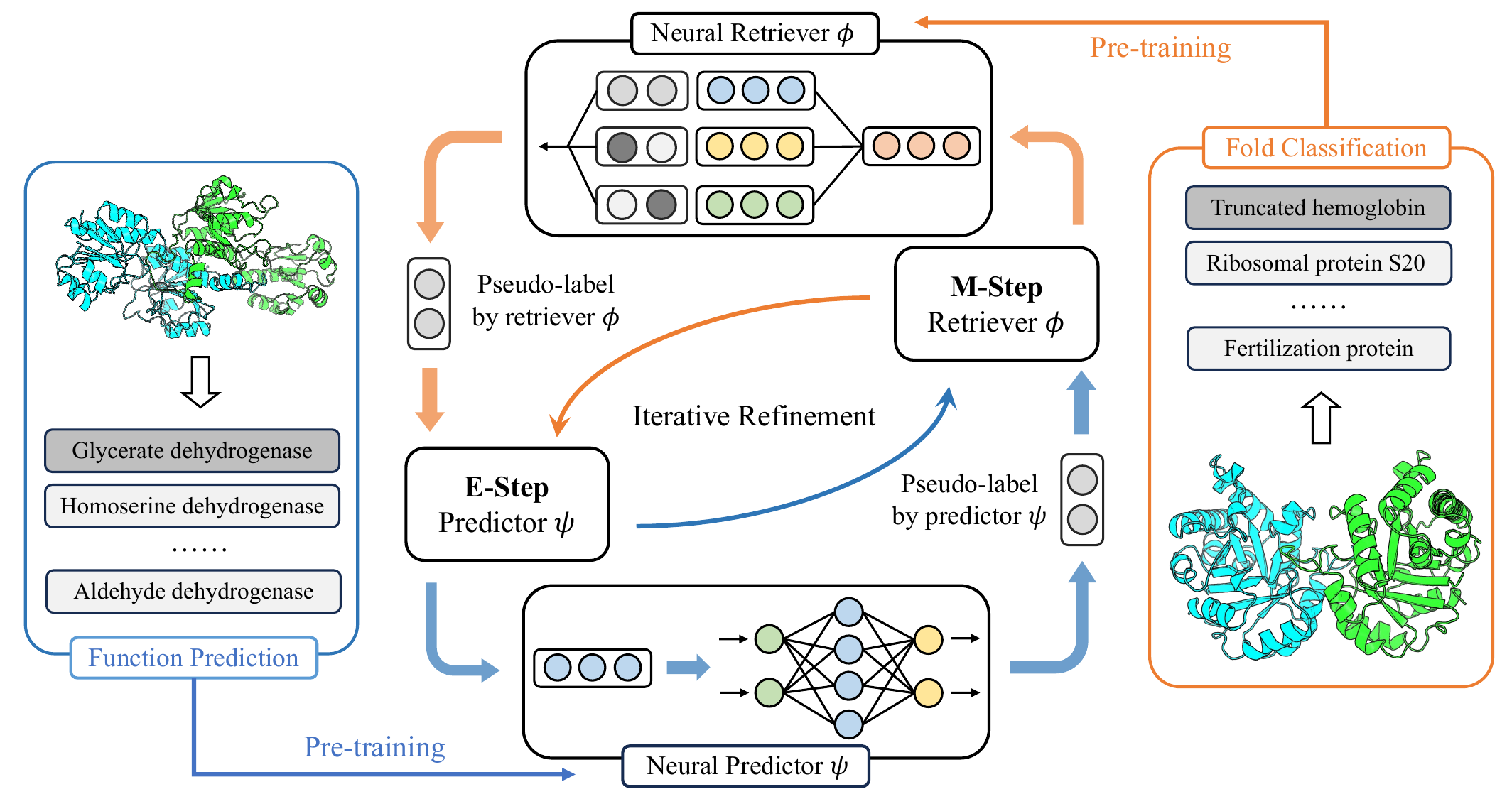}
    \caption{Overview of ProtIR. In the E-step and M-step, the neural predictor $\psi$ and retriever $\phi$ are trained, respectively, and their predictions iteratively refine each other. Before iterative refinement, the predictor $\psi$ and retriever $\phi$ are pre-trained on function prediction and fold classification, respectively.}
    \label{fig:protir}
\end{figure*}

The effectiveness of retriever-based methods highlights the importance of modeling the similarity between proteins.
Therefore, our framework is designed to model the joint distribution of observed function labels given the whole protein set, denoted as $p(\vy_L|\vx_V)$. However, directly maximizing this log-likelihood function is challenging due to the presence of unobserved protein function labels. Thus, we opt to optimize the evidence lower bound (ELBO) of the log-likelihood function instead:
\begin{equation*}
\resizebox{\linewidth}{!}{
    $p(\vy_L|\vx_V) \ge \E_{q(\vy_U|\vx_U)}[\log p(\vy_L,\vy_U|\vx_V)-\log q(\vy_U|\vx_U)],$
}
\end{equation*}
where $q(\vy_U|\vx_U)$ denotes a proposal distribution over the unobserved function labels $\vy_U$. The equality is achieved when the proposal distribution aligns with the posterior distribution, \emph{i.e.}, $q(\vy_U|\vx_U)=p(\vy_U|\vy_L,\vx_V)$.

The ELBO can be maximized through alternating optimization between the proposal distribution $q$ (E-step) and the model distribution $p$ (M-step). In the E-step, we fix the distribution $p$ and optimize the proposal distribution $q$ to minimize the KL divergence $\text{KL}(q(\vy_U|\vx_U)||p(\vy_U|\vx_V,\vy_L))$, to tighten the lower bound. 
Then, in the M-step, we keep the distribution $q$ fixed and optimize the distribution $p$ to maximize the log-likelihood function. However, direct optimization involves calculating the joint probability $p(\vy_L,\vy_U|\vx_V)$, which requires
the computationally intensive partition function in $p$. To circumvent this, we optimize the pseudo-likelihood function~\citep{besag1975statistical}:
\begin{equation}
\label{eq:pseudo}
\begin{split}
    &\begin{matrix}\E_{q(\vy_U|\vx_U)}[\log p(\vy_L,\vy_U|\vx_V)] \end{matrix}\\
    \approx& \begin{matrix}\E_{q(\vy_U|\vx_U)}[\sum_{n\in V} \log p(\vy_n|\vx_V,\vy_{V\backslash n})]\end{matrix}.
\end{split}
\end{equation}

\subsection{Parameterization}

We now discuss how to parameterize the distributions $p$ and $q$ with retrievers and predictors, respectively.
For the proposal distribution $q$, we adopt a mean-field assumption, assuming independence among function labels for different proteins. This leads to the factorization:
\begin{equation}
    q_\psi(\vy_U|\vx_U) = \begin{matrix}
        \prod_{n\in U} q_\psi(\vy_n|\vx_n),
    \end{matrix}
\end{equation}
where each term $q_\psi(\vy_n|\vx_n)$ is parameterized using an MLP head applied to the representations outputted from a protein encoder $\psi$ as introduced in Sec.~\ref{sec:def}.

On the other hand, the distribution $p(\vy_n|\vx_V,\vy_{V\backslash n})$ aims to utilize the protein set $\vx_V$ and other node labels $\vy_{V\backslash n}$ to characterize the label distribution of each protein $n$. 
This formulation aligns naturally with a retriever-based method by retrieving similar proteins from the labeled set. 
Hence, we model $p_\phi(\vy_n|\vx_V,\vy_{V\backslash n})$ with a retriever $\phi$ as in (\ref{eq:retriver}) and (\ref{eq:kernel}) to model the relationship between different proteins.
Note that the choices of the predictor $\psi$ and retriever $\phi$ are flexible and do not require the same encoder architecture.
Next, we elaborate on the optimization of both the predictor distribution $q_\psi$ and the retriever distribution $p_\phi$.

\subsection{E-Step: Predictor Optimization}
\label{sec:estep}

In the E-step, we keep the retriever $\phi$ fixed and optimize the predictor $\psi$ to maximize the evidence lower bound, allowing the retriever's understanding of global protein relationships to be distilled into the predictor. The goal is to minimize the KL divergence between the proposal distribution and the posterior distribution, expressed as $\text{KL}(q_\psi(\vy_U|\vx_U)||p_\phi(\vy_U|\vx_V,\vy_L))$.
Directly optimizing this divergence is challenging due to the reliance on the entropy of $q_\psi(\vy_U|\vx_U)$, the gradient of which is difficult to handle. To circumvent this, we adopt the wake-sleep algorithm~\citep{hinton1995wake} to minimize the reverse KL divergence, leading to the following objective to maximize:
\begin{equation}
\begin{split}
    &-\text{KL}(p_{\phi}(\vy_U|\vx_V,\vy_L)||q_{\psi}(\vy_U|\vx_U)) \\
    =\;& 
    \E_{p_{\phi}(\vy_U|\vx_V,\vy_L)}[\log q_{\psi}(\vy_U|\vx_U)] + \text{const} \\
    =\;&\begin{matrix}\sum_{n\in U}\E_{p_{\phi}(\vy_n|\vx_V,\vy_L)}[\log q_{\psi}(\vy_n|\vx_n)] + \text{const}\end{matrix},
\end{split}
\end{equation}
where const denotes the terms irrelevant with $\psi$.
This is more tractable as it avoids the need for the entropy of $q_\psi(\vy_U|\vx_U)$.
To sample from the distribution ${p_{\phi}(\vy_U|\vx_V,\vy_L)}$, we annotate the unlabeled proteins by employing $\phi$ to retrieve the most similar proteins from the labeled set using (\ref{eq:retriver}). Additionally, the labeled proteins can be used to train the predictor and prevent catastrophic forgetting~\citep{mccloskey1989catastrophic}. Combining this with the pseudo-labeling objective, we arrive at the final objective function for training the predictor:
\begin{equation}
\label{eq:eloss}
\begin{split}
        \max\nolimits_{\psi}\; & \begin{matrix}
        \sum_{n\in L} \log q_{\psi}(\vy_n|\vx_n)\end{matrix} \\
        +\;\;&\begin{matrix}\sum_{n\in U} \E_{p_{\phi}(\vy_n|\vx_V,\vy_L)}[\log q_{\psi}(\vy_n|\vx_n)].\end{matrix}
\end{split}
\end{equation}
Intuitively, the first term is a supervised training objective, and the second term acts as a knowledge distillation process, making the predictor align with the labels from the retriever.


\subsection{M-Step: Retriever Optimization}

In the M-step, our objective is to keep the predictor $\psi$ fixed and fine-tune the retriever $\phi$ to maximize the pseudo-likelihood, as introduced in (\ref{eq:pseudo}).
Similar to Sec.~\ref{sec:estep}, we sample the pseudo-labels $\hat{\vy}_U$ from the predictor distribution $q_\psi$ for unlabeled proteins $\vx_U$.
Consequently, the pseudo-likelihood objective can be reformulated as follows:
\begin{equation}
\label{eq:mstep}
\begin{split}
    &\begin{matrix}\E_{q(\vy_U|\vx_U)}[\sum_{n\in V} \log p(\vy_n|\vx_V,\vy_{V\backslash n})]\end{matrix} \\
    =\;& \begin{matrix}
    \sum_{n\in U} \log p_{\phi}(\hat{\vy}_n|\vx_V,\vy_L,\hat{\vy}_{U\backslash n}) \end{matrix} \\
    +\;& \begin{matrix}\sum_{n\in L} \log p_{\phi}(\vy_n|\vx_V,\vy_{L\backslash n},\hat{\vy}_{U}).
    \end{matrix}
\end{split}
\end{equation}
Again, the first term represents a knowledge distillation process from the predictor to the retriever via pseudo-labels, while the second is a supervised loss with observed labels.

The optimization of the retriever distribution $p_\phi$ involves learning the kernel functions $\gK(\cdot,\cdot)$ by aligning representations of proteins with identical function labels and pushing apart those with different labels.
One potential approach to the problem is supervised contrastive learning~\citep{khosla2020supervised}.
However, defining and balancing positive and negative samples in contrastive learning becomes challenging when dealing with the multiple binary labels in (\ref{eq:mstep}).
To simplify the training of the retriever $\phi$, we transform the contrastive learning into a straightforward multiple binary classification problem akin to the predictor $\psi$.
We accomplish this by introducing an MLP head over the representations outputted by $\phi$, denoted as $\tilde{p}_\phi(\vy_n|\vx_n) = \sigma(\text{MLP}(\phi(\vx_n)))$ and optimize it using binary cross entropy loss as outlined in (\ref{eq:predictor}).
Formally, the M-step can be expressed as:
\begin{equation}
\label{eq:mloss}
\resizebox{.89\linewidth}{!}{
    $\begin{matrix}
        \max_{\phi}\; \sum_{n\in U} \log \tilde{p}_{\phi}(\hat{\vy}_n|\vx_n) + \sum_{n\in L} \log \tilde{p}_{\phi}(\vy_n|\vx_n).
    \end{matrix}$
}
\end{equation}
By training the model for binary classification, proteins with similar function labels are assigned with similar representations, increasing the differences between different function classes. 
During inference, we integrate the trained retriever $\phi$ back into the original formulation in (\ref{eq:retriver}).

Finally, the workflow of the EM algorithm is summarized in Fig.~\ref{fig:protir} and Alg.~\ref{alg:inference}.
In practice, we start from a pre-trained predictor $q_\psi$ using labeled function data as in (\ref{eq:predictor}) and a retriever $p_\phi$ infused with structrual information from the fold classfication task as in (\ref{eq:fold}).
We use validation performance as a criterion for tuning hyperparameters and early stopping.
The iterative refinement process typically converges within five rounds, resulting in minimal additional training time.


\begin{algorithm}[t]
    \footnotesize
    \captionsetup{font=footnotesize}\caption{\method Algorithm}
    \textbf{Input:} Labeled proteins $\vx_L$ and their function labels $\vy_L$, unlabeled proteins $\vx_U$.\\
    \textbf{Output:} Function labels $\vy_U$ for unlabeled proteins $\vx_U$.\\
    \vspace{-1em}
    \begin{algorithmic}[1]
        \STATE{Pre-train $q_\psi$ with function labels $\vy_L$ according to (\ref{eq:predictor});}
        \STATE{Pre-train $p_\phi$ on fold classification according to (\ref{eq:fold});}
        \WHILE{not converge}
            \STATE{$\boxdot$ \emph{E-step: Predictor Learning}}
            \STATE{Annotate unlabeled proteins with $p_\phi$ and $\vy_L$ with (\ref{eq:retriver});} 
            \STATE{Update $q_\psi$ according to (\ref{eq:eloss});} 
            \STATE{$\boxdot$ \emph{M-step: Retriever Learning}}
            \STATE{Annotate unlabeled proteins with pseudo-label $\hat{\vy}_U$ from $q_\psi$;} 
            \STATE{Set $\hat{\vy}=(\vy_L,\hat{\vy}_U)$ and update $p_\phi$ with (\ref{eq:mloss});} 
        \ENDWHILE
        \STATE{Classify each unlabeled protein $\vx_n$ with $p_\psi$ and $q_\phi$}
    \end{algorithmic}
    \label{alg:inference}
\end{algorithm}
\section{Related Work}

\textbf{Protein Representation Learning.}
Previous research focuses on learning protein representations from diverse modalities, including sequences~\citep{lin2023evolutionary}, multiple sequence alignments~\citep{rao2021msa}, and structures~\citep{zhang2022protein}. 
Sequence-based methods treat protein sequences as the fundamental language of life, pre-training large models on billions of sequences~\citep{tape2019,elnaggar2020prottrans,rives2021biological}. 
Structure-based methods capture different levels of protein structures, including residue-level~\citep{gligorijevic2021structure,zhang2022protein}, atom-level structures~\citep{jing2021equivariant,hermosilla2020intrinsic}.
Diverse self-supervised learning algorithms have been developed to pre-train structure encoders, such as contrastive learning~\citep{zhang2022protein}, self-prediction~\citep{chen2022structure}, and denoising~\citep{zhang2023physics}. Recent efforts have been devoted to integrating sequence- and structure-based methods~\citep{wang2022lm,zhang2023enhancing}.

\textbf{Retriever-Based Methods.}
Retriever-based methods, starting with the k-nearest neighbors (k-NN) approach~\citep{Fix1989DiscriminatoryA,Cover1967NearestNP}, represent a critical paradigm in the field of machine learning and information retrieval, with application in text~\citep{Khandelwal2020Generalization,Borgeaud2021ImprovingLM}, image~\citep{Papernot2018DeepKN,Borgeaud2021ImprovingLM}, and video generation~\citep{Jin2023DiffusionRetGT}.
Designing protein retrievers to capture similar evolutionary and structural information has been an important topic for decades~\citep{chen2018comprehensive}.
These retrievers can be employed for function annotation~\citep{conesa2005blast2go,ma2023retrieved,yu2023enzyme}.

In this study, we take the first systematic evaluation of modern methods from both categories for function annotation. 
Moreover, we propose a novel iterative refinement framework to combine the predictor- and retriever-based methods, maximizing the utility of scarce function labels.
More related work can be found in App.~\ref{app:sec:related}.



\begin{figure*}[t]
    \centering
    \includegraphics[width=\linewidth]{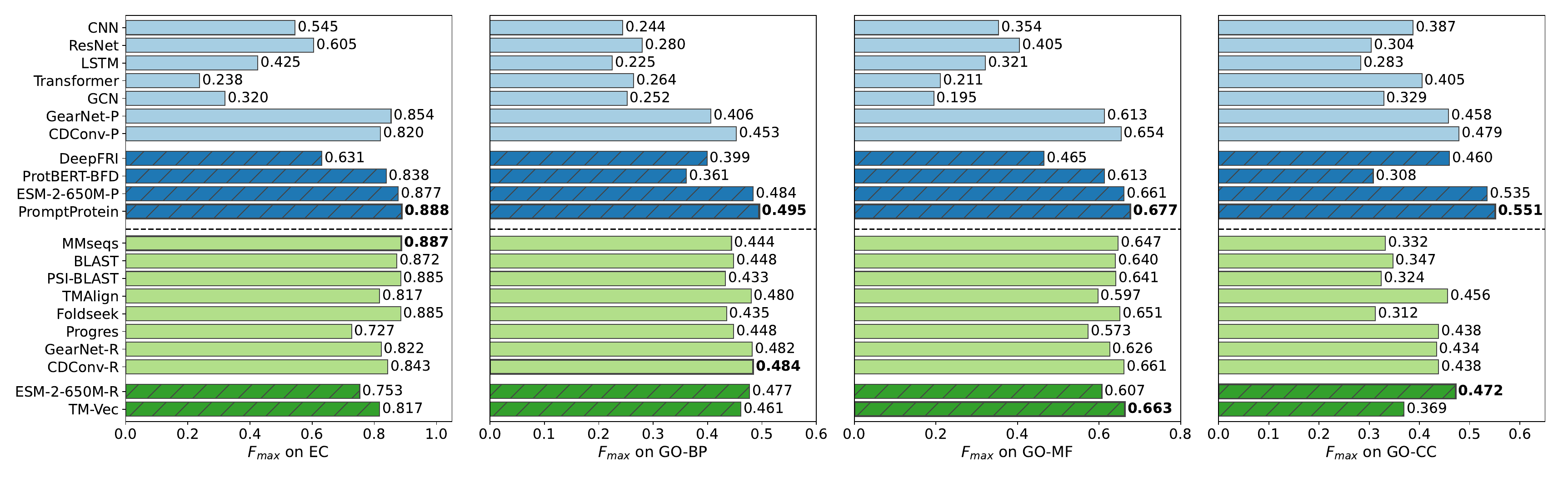}
    \vspace{-2em}
    \caption{F\textsubscript{max} on four protein function annotation tasks under the 95\% cutoff. 
    The upper section (in blue) displays the results for methods based on predictors, whereas the lower section (in green) presents results for methods based on retrievers. The best methods in each section are highlighted in bold. Methods employing protein language models as predictors and retrievers are highlighted with hatches in dark blue and dark green, respectively.
    Detailed results are shown in Tab.~\ref{tab:all_result}.}
    \label{fig:benchmark}
\end{figure*}

\section{Experiments}
\label{sec:exp}
In this section, we address two main research questions: how predictor- and retriever-based methods perform on function annotation, and how retriever-based insights can enhance predictor-based methods. 
To tackle these questions, we focus our study on function annotation tasks (see Sec.~\ref{sec:setup}). 
For the first question, we benchmark standard baselines from both categories of methods (Sec.~\ref{sec:exp:benchmark}). 
For the second, we explore incorporating inter-protein similarity in predictors, by applying the ProtIR framework to predictors without massive pre-training (Sec.~\ref{sec:exp:protir}).


\subsection{Experimental Setup}
\label{sec:setup}

We evaluate the methods using two function annotation tasks in~\citet{gligorijevic2021structure}.
The first task, \textbf{Enzyme Commission (EC) prediction}, involves predicting the EC numbers for proteins, indicating their role in biochemical reactions, focusing on the third and fourth levels of the EC tree~\citep{webb1992enzyme}.
The second task, \textbf{Gene Ontology (GO) prediction}, determines if a protein is associated with specific GO terms, classifying them into molecular function (MF), biological process (BP), and cellular component (CC) categories, each reflecting different aspects of protein function.
More details can be found in App.~\ref{app:exp:dataset}.

To ensure a rigorous evaluation, we follow the multi-cutoff split methods outlined in~\citet{gligorijevic2021structure}. Specifically, we ensure that the test sets only contain PDB chains with a sequence identity of no more than 30\%, 50\%,  and 95\% to the training set, aligning with the approach used in~\citet{wang2022lm}. The evaluation of performance is based on the protein-centric maximum F-score, denoted as F\textsubscript{max}, a commonly used metric in the CAFA challenges~\citep{radivojac2013large}.

\begin{table*}[t]
    \centering
    \caption{F\textsubscript{max} on EC and GO prediction with iterative refinement and transductive learning baselines.
    }
    \vspace{-0.8em}
    \label{tab:framework}
    \begin{threeparttable}
    \begin{adjustbox}{max width=\linewidth}
        \begin{tabular}{l|ccccccccccccccccc}
            \toprule[2pt]
            \multirow{2}{*}{\large{\bf{Model}}} & \multirow{2}{*}{\large{\bf{Method}}} &
            \multicolumn{3}{c}{\bf{EC}}&&
            \multicolumn{3}{c}{\bf{GO-BP}} && 
            \multicolumn{3}{c}{\bf{GO-MF}} && 
            \multicolumn{3}{c}{\bf{GO-CC}}
            \\
            \cmidrule{3-5}
            \cmidrule{7-9}
            \cmidrule{11-13}
            \cmidrule{15-17}
            & & 30\% & 50\% & 95\% &&
            30\% & 50\% & 95\% && 
            30\% & 50\% & 95\% && 
            30\% & 50\% & 95\% 
            \\
            \midrule[1.5pt]
            \multirow{6}{*}{\bf{GearNet}} & {Predictor} & 0.700 & 0.769 & 0.854 && 0.348 & 0.359 & 0.406 && 0.482 & 0.525 & 0.613 && 0.407 & 0.418 & 0.458 \\
            \cmidrule{2-17}
            & {Pseudo-labeling} & 0.699 & 0.767 & 0.852 && 0.344 & 0.355 & 0.403 && 0.490 & 0.532 & 0.617 && 0.420 & 0.427 & 0.466 \\
            & {Temporal ensemble} & 0.698 & 0.765 & 0.850 && 0.339 & 0.348 & 0.399 && 0.480 & 0.526 & 0.613 && 0.402 & 0.412 & 0.454 \\
            & {Graph conv network} & 0.658 & 0.732 & 0.817 && 0.379 & 0.395 & 0.443 && 0.479 & 0.528 & 0.609 && 0.437 & 0.452 & 0.483 \\
            \cmidrule{2-17}
            & \bf{\method} & \bf{0.743} & \bf{0.810} & \bf{0.881} && \textcolor{blue}{\bf{0.409}} & \textcolor{red}{\bf{0.431}} &	\textcolor{red}{\bf{0.488}}&& \bf{0.518} & \bf{0.564} &	\bf{0.650}&& \bf{0.439} & \bf{0.452} &	\bf{0.501} \\
            & \bf{Improvement $\uparrow$} & \bf 6.1\% & \bf 5.3\% & \bf 3.1\% && \textcolor{blue}{\bf{17.5\%}} & \textcolor{red}{\bf{20.0\%}} &	\textcolor{red}{\bf{20.1\%}} && \bf 7.4\% & \bf 7.4\% & \bf 6.0\% && \bf 7.8\% & \bf 8.1\% &	\bf 9.3\% \\
            \midrule[1.5pt]
            \multirow{6}{*}{\bf{CDConv}} & {Predictor} & 0.634 & 0.702 & 0.820 && 0.381 & 0.401 & 0.453 && 0.533 & 0.577 & 0.654 && 0.428 & 0.440 & 0.479 \\
            \cmidrule{2-17}
            & {Pseudo-labeling} & 0.722 & 0.784 & 0.861 && 0.397 & 0.413 & 0.465 && 0.529 & 0.573 & 0.653 && 0.445 & 0.458 & 0.495 \\
            & {Temporal ensemble} & 0.721 & 0.785 & 0.862 && 0.381 & 0.394 & 0.446 && 0.523 & 0.567 & 0.647 && 0.444 & 0.455 & 0.492 \\
            & {Graph conv network} & 0.673 & 0.742 & 0.818 && 0.380 & 0.399 & 0.455 && 0.496 & 0.545 & 0.627 && 0.417 & 0.429 & 0.465 \\
            \cmidrule{2-17}
            & \bf{\method} & \textcolor{red}{\bf{0.769}} & \textcolor{blue}{\bf{0.820}} & \bf{0.885} && \textcolor{blue}{\bf{0.434}} & \textcolor{blue}{\bf{0.453}} &	\textcolor{red}{\bf{0.503}}&& \bf{0.567} & \bf{0.608} &	\bf{0.678}&& \bf{0.447} & \bf{0.460} &	\bf{0.499} \\
            & \bf{Improvement $\uparrow$} & \textcolor{red}{\bf{21.2\%}} & \textcolor{blue}{\bf{16.8\%}} & \bf 4.2\% && \textcolor{blue}{\bf{13.9\%}} & \textcolor{blue}{\bf{12.9\%}} & \textcolor{red}{\bf{23.8\%}} && \bf 6.3\% & \bf 5.3\% & \bf 3.6\% && \bf 4.4\% & \bf 4.5\% &	\bf 4.1\% \\
            \midrule[1.5pt]
            \multicolumn{2}{c}{{PromptProtein}} & {0.765} & {0.823} & {0.888} && {0.439} & {0.453} & {0.495} && {0.577} & {0.600} & {0.677} && {0.532} & {0.533} & {0.551} \\
            \bottomrule[2pt]
        \end{tabular}
    \end{adjustbox}
    \begin{tablenotes}
        \item[*] \footnotesize \textcolor{red}{\bf{Red}}: over 20\% improvement; \textcolor{blue}{\bf{blue}}: 10\%-20\% improvement; \textbf{bold}: 3\%-10\% improvement.
    \end{tablenotes}
    \end{threeparttable}
\end{table*}

\subsection{Benchmark Results of Predictor- and Retriever-Based Methods}
\label{sec:exp:benchmark}

\textbf{Baselines.}
We select two categories of predictor-based baselines for comparison: (1) \emph{Protein Encoders without Pre-training}: This category includes four sequence-based encoders (CNN, ResNet, LSTM and Transformer~\citep{tape2019}) and three structure-based encoders (GCN~\citep{kipf2017semi}, GearNet~\citep{zhang2022protein}, CDConv~\citep{fan2023continuousdiscrete}).
(2) \emph{Protein Encoders with Massive Pre-training}: This includes methods based on protein language models (PLM) pre-trained on millions to billions of protein sequences, such as DeepFRI~\citep{gligorijevic2021structure}, ProtBERT-BFD~\citep{elnaggar2020prottrans}, ESM-2~\citep{lin2023evolutionary} and PromptProtein~\citep{wang2023multilevel}.
Due to computational constraints, we use ESM-2-650M for the benchmark.
We use the suffix "-P" to distinguish with methods used as retrievers.

For retriever-based methods, we considered retrievers with and without protein language models.
For those without PLMs, we select three sequence retrievers, MMSeqs~\citep{steinegger2017mmseqs2}, BLAST~\citep{altschul1990basic} and PSI-BLAST~\citep{altschul1997gapped}, and three structure retrievers, TMAlign~\citep{zhang2005tm}, Foldseek~\citep{van2023fast} and Progres~\citep{greener2022fast}.
Additionally, we train two neural structure retrievers by using GearNet and CDConv on fold classification tasks as in (\ref{eq:fold}), denoted as GearNet-R and CDConv-R, respectively.
For retrievers with PLMs, we consider using ESM-2-650M~\citep{lin2023evolutionary}, denoted as ESM-2-650M-R, and recently proposed TM-Vec~\citep{Hamamsy2023ProteinRH} for retrieving similar proteins.

\textbf{Training.}
For predictor-based methods, except for GearNet and ESM-2-650M, all results were obtained from a previous benchmark~\citep{zhang2022protein}. We re-implement GearNet, optimizing it following CDConv's implementation with a 500-epoch training, leading to significant improvements over the original paper~\citep{zhang2022protein}.
For ESM-2-650M, we fine-tune the model for 50 epochs. 
For GearNet and CDConv retrievers, we train them on the fold dataset for 500 epochs, selecting the checkpoint with the best validation performance as the final retrievers.
Detailed training setup for other retriever-based methods is provided in App.~\ref{app:exp:setup}.  All these models are trained on one A100 GPU.

\begin{figure*}[t]
    \centering
    \includegraphics[width=0.95\linewidth]{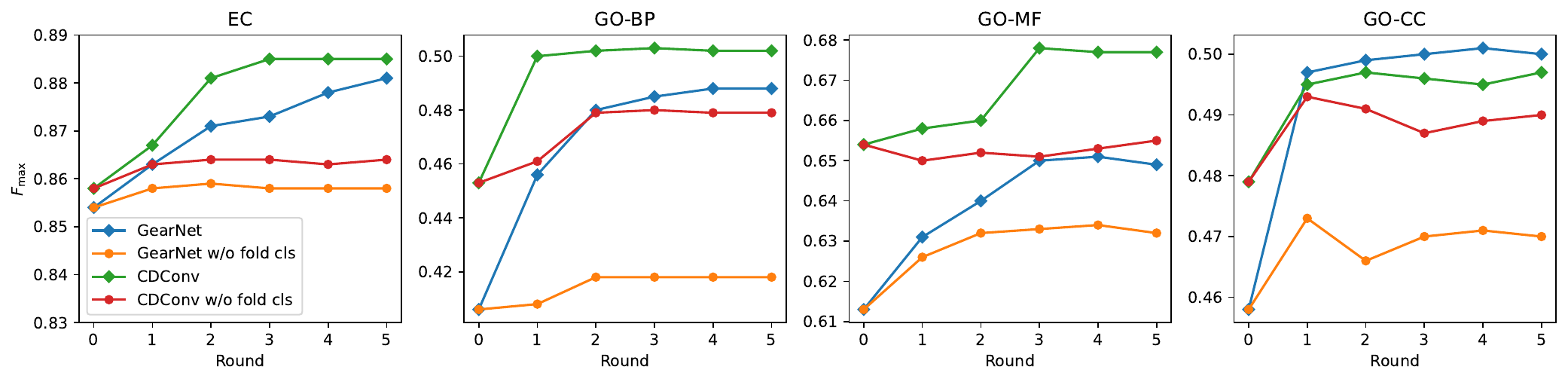}
    \vspace{-1em}
    \caption{F\textsubscript{max} on function annotation tasks \emph{vs.} number of rounds in iterative refinement. Besides default \method, we also depict curves without retriever pre-training on fold classification, highlighting the impact of injecting structural information.}
    \label{fig:round}
\end{figure*}

\textbf{Results.}
The benchmark results are presented in Fig.~\ref{fig:benchmark} and Tab.~\ref{tab:all_result} (App.~\ref{app:sec:benchmark}). 
Here is an analysis of the findings\footnote{Notably, the results for GO-CC differ from other tasks. GO-CC aims to predict the cellular compartment, which is less directly related to the protein's function itself.}:\\[0.3em]
\emph{Firstly, retriever-based methods exhibit comparable or superior performance to predictor-based methods without pre-training.} A comparison between methods in light blue and green in Fig.~\ref{fig:benchmark} reveals that retrievers can outperform predictors even without training on function labels. This supports the hypothesis that proteins sharing evolutionary or structural information have similar functions.\\[0.3em]
\emph{Secondly, predictor-based methods using fine-tuned Protein Language Models (PLMs) significantly outperform retrievers.} 
By comparing the results in dark blue and green, it can be observed that PLM-based predictors, \emph{e.g.}, ESM-2 and PromptProtein excels on all consider functions.
This aligns with the principle that deep learning techniques effectively leverage large pre-training datasets, enabling neural predictors to capture more evolutionary information than traditional, hard-coded retrievers.\\[0.3em]
\emph{Thirdly, contrary to expectations, structure retrievers do not always outperform sequence retrievers.} As shown in the light green part of the figure, sequence retrievers like MMSeqs perform better than structure retrievers like CDConv-R on the EC task but are less effective for the GO tasks. This discrepancy may underscore the importance of evolutionary information for enzyme catalysis, while structural aspects are more crucial for molecular functions.\\[0.3em]
\emph{Fourthly, a universal retriever excelling across all functions is still lacking.} For instance, the best structure retriever, CDConv-R, underperforms in EC number predictions, whereas sequence retrievers struggle with GO predictions. This suggests that different functions rely on varying factors, which may not be fully captured by these general-purpose sequence and structure retrievers.\\[0.5em]
To further show the potential of retriever-based methods, we perform experiments on real-world settings for EC number annotation in App.~\ref{app:exp:clean}, where retriever-based methods without function-specific training can outperform EC annotation tools. 
To summarize, while retriever-based methods demonstrate potential for accurate function annotation without massive pre-training, a universal retriever with state-of-the-art performance across all functions is yet to be developed. Nonetheless, the concept of inter-protein similarity modeling shows potential for enhancing function annotation accuracy.


\subsection{Results of Iterative Refinement Framework}
\label{sec:exp:protir}


\textbf{Setup.}
We employ the GearNet and CDConv trained on EC and GO as our backbone models and conduct a comprehensive evaluation by comparing our proposed iterative refinement framework with several baseline methods.
As the iterative refinement framework falls under the category of transductive learning~\citep{Vapnik2000TheNO}, we benchmark our approach against three well-established deep semi-supervised learning baselines: pseudo-labeling~\citep{Lee2013PseudoLabelT}, temporal ensemble~\citep{laine2017temporal}, and graph convolutional networks~\citep{kipf2017semi}. These baselines are trained for 50 epochs.
For our method, we iterate the refinement process for 5 rounds, halting when no improvements are observed on the validation set. In each iteration, both E-step and M-step are trained for 30 epochs.

\textbf{Results.}
The results are summarized in Tab.~\ref{tab:framework}. Notably, our proposed iterative refinement consistently demonstrates substantial improvements across various tasks and different backbone models when compared to both vanilla predictors and other transductive learning baselines. On average, GearNet showcases a remarkable improvement of 9.84\%, while CDConv exhibits an impressive 10.08\% enhancement, underscoring the effectiveness of our approach.
Moreover, in comparison to the state-of-the-art predictor-based method, PromptProtein, CDConv achieves similar performance on EC, GO-BP, and GO-MF tasks while reducing the need for pre-training on millions of sequences. 

\begin{figure}[t]
    \centering
    \includegraphics[width=0.9\linewidth]{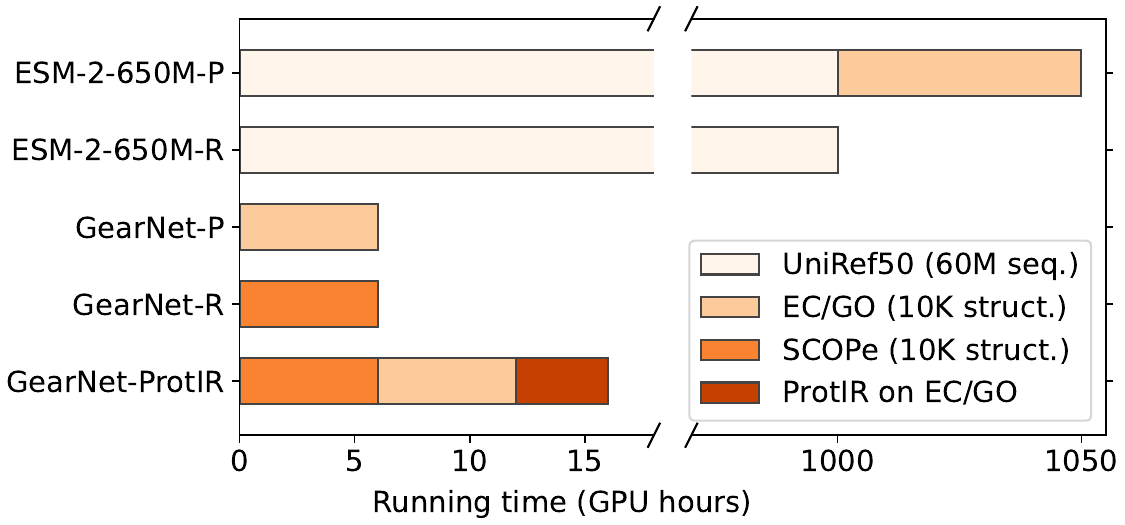}
    \caption{Training time of different methods at pre-training and fine-tuning stages on A100, where pre-training protein language models takes over 1K GPU hours.}
    \label{fig:time}
    \vspace{-1em}
\end{figure}

\textbf{Time Analysis.}
To give a clear understanding of the efficiency of \method, we list the training times for various function annotation methods, both with and without protein language models (PLMs), in Fig.~\ref{fig:time}.
Notably, since the inference time for all methods typically does not exceed 1 GPU hour, we exclude it from our comparison.
The table indicates that PLM-based methods, such as ESM-2-650M, often require massive pre-training, involving thousands of hours on millions of protein sequences.
Also, fine-tuning these large models requires more time than small structure-based encoders (17M params. in GearNet \emph{v.s.} 650M params. in ESM-2-650M), making it difficult to scaling to large downstream sets with limited resources.
In contrast, methods utilizing the \method framework can attain comparable performance levels without time-consuming pre-training phases.
These methods, by merely pre-training on datasets of the order of tens of thousands and applying iterative refinement in downstream tasks, demonstrate competitive performance against PLM-based approaches. 
This analysis underscores the efficiency and effectiveness of the \method framework in protein function annotation.

\textbf{Analysis and Ablation Study.}
To analyze the iterative refinement process, we present the test performance curve against the number of rounds in Fig.~\ref{fig:round}. The results reveal a consistent enhancement in test performance for both models, with convergence typically occurring within five rounds. This underscores the efficiency of our iterative framework in yielding performance gains relatively swiftly. 
Additionally, we examine the impact of injecting structural information into the retriever by comparing results with and without fold classification pre-training. Notably, while improvements are observed without fold pre-training, the performance is significantly superior with this pre-training, emphasizing the importance of incorporating structural insights.

\begin{figure*}[t]
    \centering
    \includegraphics[width=\linewidth]{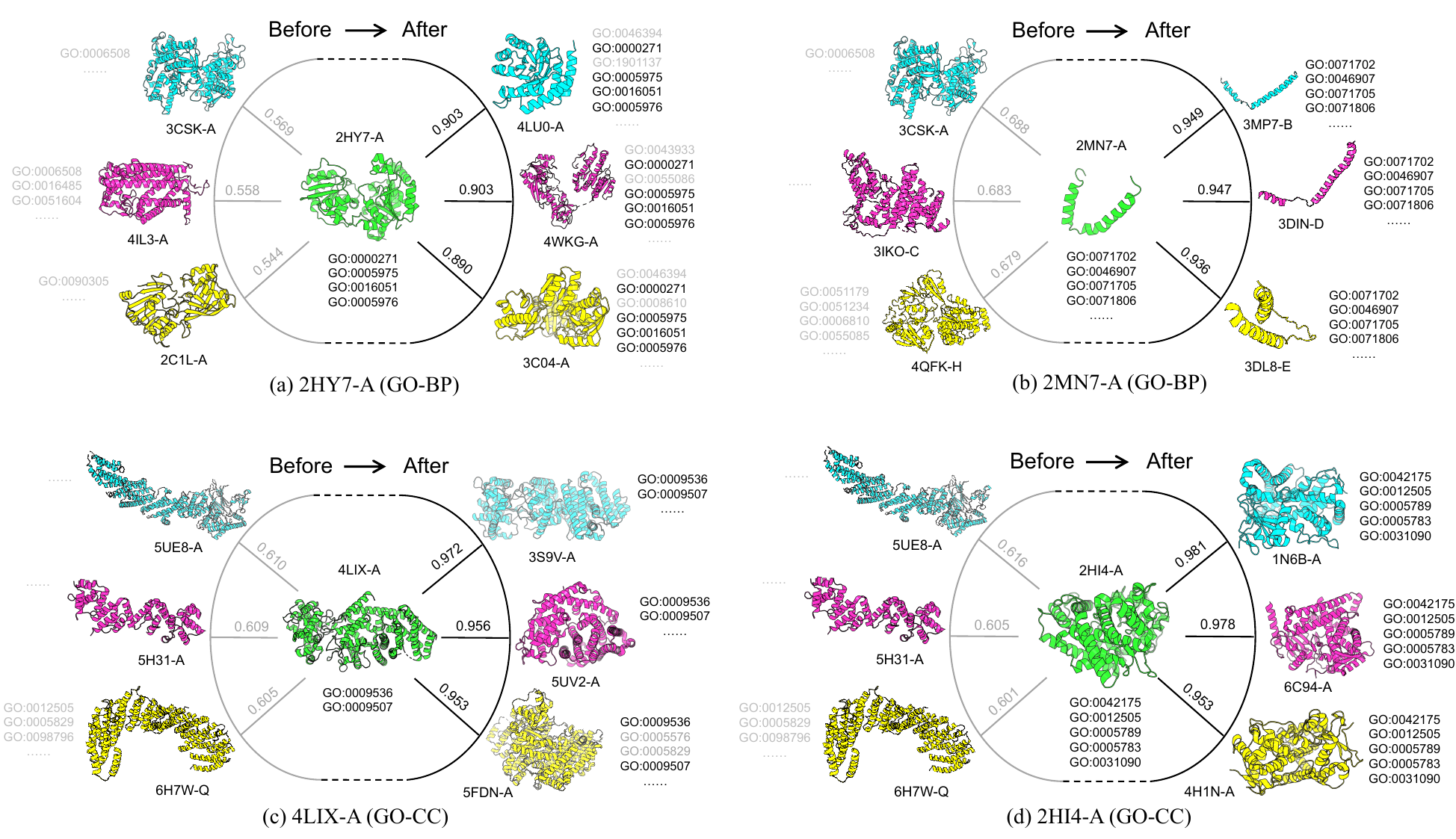}
    \caption{Interpretability analysis for GearNet-P's prediction before and after applying \method. For GO-BP and GO-CC, we select two representative samples and identify their top-3 closest neighbors in GearNet-P's latent space in terms of cosine similarity. The results for GearNet-P before \method are shown in grey, whereas the results of GearNet-P after \method are shown in black. The corresponding cosine similarities are listed on the figure.
    The analysis clearly shows that \method significantly improves the similarity between representations of proteins with similar functions, even without explicit supervision. This finding highlights the efficacy of employing retrievers as a means to refine predictors.}
    \label{fig:interpret}
\end{figure*}

\textbf{Interpretability.}
One key advantage of retriever-based methods lies in their strong interpretability, particularly in terms of presenting the most closely related proteins as reference examples. 
Now we investigate whether predictor-based methods can adopt this merit from retriever-based methods during the iterative refinement process in \method. 
For this purpose, we focus on two predictors, specifically GearNet applied to GO-BP and GO-CC, which exhibit significant improvements after applying \method. 
We select two representative samples from each task's test set. 
Subsequently, we explore the training set to find the top three closest neighboring proteins, based on cosine similarity between representations generated by GearNet-P, both before and after the \method application. 
The results are illustrated in Fig.~\ref{fig:interpret}. 
The figure reveals that, for all four analyzed samples, GearNet-P initially shows moderate similarity for the nearest neighbors, with cosine similarities ranging between 0.5 and 0.7. 
This suggests that the model is initially uncertain about which proteins in the training set perform similar functions. 
However, a notable improvement is observed post \method application, where GearNet-P effectively aligns representations for proteins with similar functions. This result underscores the ability of \method to transfer inter-protein similarity knowledge from retrievers to predictors, achieving this without any direct supervision on pairwise similarity.

\vspace{-0.5em}
\section{Conclusion}

In this study, we conducted a comprehensive evaluation of various sequence and structure retriever-based methods, as well as predictor-based approaches, in the context of protein function annotation tasks. Our well-motivated \method framework combines the strengths of both methodologies, resulting in significant enhancements over vanilla predictor-based methods. 
However, our research was constrained by resource limitations, leading us to apply the \method framework only to smaller structure encoders and thus failing to achieve the state-of-the-art performance. Future studies should explore its application with cutting-edge PLM-based methods. 
Additionally,
further exploration into other applications besides function annotation should be considered, such as protein engineering.




\section*{Acknowledgments}

The authors would like to thank Minghao Xu, Huiyu Cai and Bozitao Zhong for their helpful discussions and comments.

This project is supported by AIHN IBM-MILA partnership program, the Natural Sciences and Engineering Research Council (NSERC) Discovery Grant, the Canada CIFAR AI Chair Program, collaboration grants between Microsoft Research and Mila, Samsung Electronics Co., Ltd., Amazon Faculty Research Award, Tencent AI Lab Rhino-Bird Gift Fund, a NRC Collaborative R\&D Project (AI4D-CORE-06) as well as the IVADO Fundamental Research Project grant PRF-2019-3583139727.

\bibliography{reference}
\bibliographystyle{icml2024}

\newpage
\appendix
\onecolumn

\section{More Related Work and Broader Impact}
\label{app:sec:related}

\textbf{Protein Retriever.}
In the domain of proteins, retriever-based methods have long been employed for function annotation~\citep{conesa2005blast2go}, utilizing both sequence~\citep{altschul1990basic,melvin2011detecting,buchfink2021sensitive,Hamamsy2023ProteinRH} and structure-based approaches~\citep{Shindyalov1998ProteinSA,Yang2006ProteinSD,Zhao2013SSWLA,Holm2019BenchmarkingFD,trinquier2022swampnn,greener2022fast,van2023fast}.
Recent endeavors have extended retrievers to retrieve similar sequences from expansive databases, augmenting inputs and subsequently enhancing function annotation performance~\citep{ma2023retrieved,zou2023antibody,dickson2023fine,kilinc2023improved,chen2023learning}. 
Instead of designing a new protein retriever, our work proposes a general strategy to train a neural structure retriever and studies how to use the idea of inter-protein similarity modeling to improve function annotation accruacy.

\textbf{Protein Network Propagation for Function Prediction.}
Besides directly measuring inter-protein similarities based on sequences and structures, there is a parallel line of research that focuses on function annotation through protein-protein interaction (PPI) networks, exemplified by tools like STRING~\citep{szklarczyk2019string}. These networks map both direct physical and indirect functional interactions among proteins.
Recent approaches in this domain involve functional label propagation within these networks~\citep{mostafavi2008genemania,wang2017prosnet,you2019netgo,cho2016compact,kulmanov2018deepgo,yao2021netgo}, and adapting these methods to PPI networks of newly sequenced species~\citep{you2021deepgraphgo,torres2021protein}. However, a key limitation of these methods is that they are not able to make predictions for newly sequenced proteins absent in existing PPI networks.
Moreover, knowing protein-protein interactions is essentially a more difficult challenge, as it requires a more comprehensive understanding of protein properties. 
These problems make this line of work hard to use in real-world settings.

\textbf{Transductive Learning.}
Our iterative refinement framework falls into the category of transductive learning, focusing on optimizing performance for specific sets of interest rather than reasoning general rules applicable to any test cases~\citep{Vapnik2006EstimationOD}.
Typical transductive learning methods encompass generative techniques~\citep{springenberg2015unsupervised,kingma2014semi}, consistency regularization approaches~\citep{rasmus2015semi,laine2017temporal}, graph-based algorithms~\citep{kipf2017semi,gilmer2017neural}, pseudo-labeling strategies~\citep{Lee2013PseudoLabelT}, and hybrid methodologies~\citep{verma2022interpolation,berthelot2019mixmatch}.
In contrast to existing approaches, our work develops a novel iterative refinement framework for mutual enhancement between predictors and retrievers.

\textbf{Broader Impact and Ethical Considerations.}
The main objective of this project is to enable more accurate protein function annotation by modeling inter-protein similarity.
Unlike traditional protein retrievers, our approach utilizes structural information in the SCOPe dataset to build a neural structure retriever.
This advantage allows for more comprehensive analysis of protein research and holds potential benefits for various real-world applications, including protein engineering, sequence and structure design.
It is important to acknowledge that powerful function annotation models can potentially be misused for harmful purposes, such as the design of dangerous drugs. We anticipate that future studies will address and mitigate these concerns.

\textbf{Limitations.}
In this study, we explore the design of a general neural structure retriever and conduct benchmarks on existing retrievers and predictors for function annotation. However, given the extensive history of protein retriever development in the bioinformatics field, it is impractical to include every retriever in our benchmark. We have chosen baselines that are typical and widely recognized within the community, acknowledging that the investigation of other promising retrievers remains a task for future research.
Our focus in this work is strictly on the application of retrievers for function annotation tasks. However, it is crucial to consider other downstream applications in future studies. For instance, protein engineering tasks, where the goal is to annotate proteins with minor sequence variations, present an important area for application.
Another limitation of our current approach is the exclusive use of the ProtIR framework with the encoder, without integrating protein language models, primarily to minimize computational expenses. Exploring the integration of this framework with larger models could yield significant insights and advancements in the field.



\section{Dataset Details}
\label{app:exp:dataset}

\begin{table}[!h]
    \centering
    \caption{Dataset statistics.}
    \label{tab:dataset}
    \begin{adjustbox}{max width=0.6\linewidth}
        \begin{tabular}{lccc}
            \toprule[2pt]
            \multirow{2}{*}{\bf{Dataset}} &
            \multicolumn{3}{c}{\bf{\# Proteins}}\\
            & \bf{\# Train} & \bf{\# Validation} & \bf{\# 30\% Test / \# 50\% Test / \# 95\% Test}\\
            \midrule[1pt]
            \bf{Enzyme Commission} & 15,550 & 1,729 & 720 / 1,117 / 1,919 \\
            \bf{Gene Ontology} & 29,898 & 3,322 & 1,717 / 2,199 / 3,416 \\
            \bf{Fold Classification} & 12,312 & - & - \\
            \bottomrule[2pt]
        \end{tabular}
    \end{adjustbox}
\end{table}

Dataset statistics are summarized in Tab.~\ref{tab:dataset}.
Details are introduced as follows.

For evaluation, we adopt two standard function annotation tasks as in previous works~\citep{gligorijevic2021structure,zhang2022protein}.
The first task, Enzyme Commission (EC) number prediction, involves forecasting the EC numbers for proteins, categorizing their role in catalyzing biochemical reactions.
We focus on the third and fourth levels of the EC hierarchy~\citep{webb1992enzyme}, forming 538 distinct binary classification challenges.
The second task, Gene Ontology (GO) term prediction, targets the identification of protein associations with specific GO terms. We select GO terms that have a training sample size between 50 and 5000.
These terms are part of a classification that organizes proteins into functionally related groups within three ontological categories: molecular function (MF), biological process (BP), and cellular component (CC).

To construct a non-redundant dataset, all PDB chains are clustered, setting a 95\% sequence identity threshold. From each cluster, a representative PDB chain is chosen based on two criteria: annotation presence (at least one GO term from any of the three ontologies) and high-quality structural resolution.
The non-redundant sets are divided into training, validation and test sets with approximate ratios 80/10/10\%.
The test set exclusively contains experimentally verified PDB structures and annotations.
We ensure that these PDB chains exhibit a varied sequence identity spectrum relative to the training set, specifically at 30\%, 50\%, and 95\% sequence identity levels. Moreover, each PDB chain in the test set is guaranteed to have at least one experimentally validated GO term from each GO category.

For pre-training a protein structure retriever, we adopt the fold classfication task~\citep{hou2018deepsf}, which is relevant with analyzing the relationship between protein structure and function, as well as in the exploration of protein evolution~\citep{hou2018deepsf}. This classification groups proteins based on the similarity of their secondary structures, their spatial orientations, and the sequence of their connections. The task requires predicting the fold class to which a given protein belongs.

For the training of our model, we utilize the main dataset obtained from the SCOP 1.75 database, which includes genetically distinct domain sequence subsets that share less than 95\% identity, updated in 2009~\citep{murzin1995scop}. This dataset encompasses 12,312 proteins sorted into 1,195 unique folds. The distribution of proteins across these folds is highly skewed: about 5\% of the folds (61 out of 1,195) contain more than 50 proteins each; 26\% (314 out of 1,195) have between 6 to 50 proteins each; and the majority, 69\% (820 out of 1,195), consist of 5 or fewer proteins per fold. The sequence lengths of the proteins in these folds vary, ranging from 9 to 1,419 residues, with most falling within the 9 to 600 range.

\section{Implementation Details}
\label{app:exp:setup}

In this subsection, we describe implementation details of retriever-based baselines and our methods.

\paragraph{BLAST and PSI-BLAST.}
We obtained the BLAST+ Version 2.14.0~\citep{altschul1990basic, camacho2009blast+} as its command line application to retrieve similar sequences for proteins in test set. For each task, we firstly built a BLAST database using \verb|-dtype prot| (indicating "protein" sequences) for the training sequences. We then searched against the database for similar sequences using  \verb|blastp| to query with parameter \verb|-evalue 0.01| and  \verb|psiblast| to query with parameters \verb|-num_iterations 4 -comp_based_stats 1 -evalue 0.01|. For the \verb|psiblast|, the hits (retrieved sequences) from the final round is adopted. The alignment score is used to rank the retrieved proteins.

\paragraph{MMSeqs.} 
We ran the MMSeqs2~\citep{steinegger2017mmseqs2} as another sequence-based retriever. The sequence database was built for both training set and test set using \verb|mmseqs createdb| command and the alignment results were obtained by searching the test database against the training database with \verb|mmseqs search| with the default configuration: \verb|-s 5.7 -e 0.001 --max-seqs 300|. Finally, the alignment results were converted into readable table using the \verb|mmseqs convertalis| and the (alignment) bit score was used to rank the retrieved records.

\paragraph{TMAlign.} 
TM-align~\citep{zhang2005tm} is a pairwise structure alignment tools for proteins based on TM-score. TM-score is a normalized similarity score in $(0,1]$ and can be used to rank the retrieved results. We ran the TM-align by enumerating all pairs between test set and training set, which forms a complete bipartite graph. Due to the intensive computational overhead, we executed the alignment with the flag \verb|-fast| and then rank the results using TM-score.

\paragraph{Foldseek.} 
Foldseek~\citep{van2023fast} is run to obtain structure-based retrieved results. We created a Foldseek database for all structures in the training set using \verb|foldseek createdb| and created search index with \verb|foldseek createindex|. Then we searched for each structure in test set against the training database using command \verb|foldseek easy-search|. All commands above were executed using 3Di+AA Gotoh-Smith-Waterman (\verb|--alignment-type 2|) with the default parameters: \verb|-s 9.5 --max-seqs 1000 -e 0.001 -c 0.0| and the alignment bit scores are used for ranking.

\paragraph{Progres.} 
Progres~\citep{greener2022fast} is a structure-based protein retrieval method based on a neural graph encoder. Firstly, we downloaded the code from the official repository as well as the trained model weights. Then we computed the graph embeddings for all the protein structures in both training and test set and all-vs-all pairwise similarity scores between them. The similarity score, as defined by \citet{greener2022fast}, is a normalized version of cosine similarity or formally $(\vv_1\cdot \vv_2/\|\vv_1\| \|\vv_2\| +1) / 2$. The similarity scores are used for ranking.

\paragraph{TM-Vec.} 
TM-vec~\citep{hamamsy2022tm} is a neural sequence alignments tool that leverages structure-base similarity data in protein databases for training. To search the retrieved results between test and training set, we downloaded and ran the codes from its official repository. Specifically, we downloaded the pretrained weights for encoders named as \verb|tm_vec_cath_model_large.ckpt|. We then built up the search database for the protein sequences in training set by running \verb|tmvec-build-database| and \verb|build-fasta-index| with default parameters. Finally, the search was performed against the database above by setting query as test set with \verb|--k-nearest-neighbors 10|. The predicted TM-score from the model is used for ranking.

For all retriever-based methods, we choose the top-$\{1,3,5,10\}$ similar proteins from the training set and tune the temperature $\tau\in\{0.03,0.1,1,10,100\}$ according to the performance on validation sets.
For neural methods, we use a batch size of $8$ and an SGD optimizer with learning rate 1e-3, weight decay 5e-4 and momentum $0.9$ for training.
The models will be trained for 500 epochs and the learning rate will decay to one tenth at the 300-th and 400-th epoch.
The code is implemented with TorchDrug~\citep{zhu2022torchdrug}.
Other training details have been introduced in Sec.~\ref{sec:exp}.

\section{Additional Experiments}

\begin{table*}[t]
    \centering
    \caption{F\textsubscript{max} on EC and GO prediction with predictor- and retriever-based methods.
    }
    \vspace{-0.8em}
    \label{tab:all_result}
    \begin{threeparttable}
    \begin{adjustbox}{max width=\linewidth}
        \begin{tabular}{lccccccccccccccccc}
            \toprule[2pt]
            & \multirow{2}{*}{{\bf{Method}}} & \multirow{2}{*}{{\bf{PLM}}} &
            \multicolumn{3}{c}{\bf{EC}}&&
            \multicolumn{3}{c}{\bf{GO-BP}} && 
            \multicolumn{3}{c}{\bf{GO-MF}} && 
            \multicolumn{3}{c}{\bf{GO-CC}}
            \\
            \cmidrule{4-6}
            \cmidrule{8-10}
            \cmidrule{12-14}
            \cmidrule{16-18}
            & & & 30\% & 50\% & 95\% &&
            30\% & 50\% & 95\% && 
            30\% & 50\% & 95\% && 
            30\% & 50\% & 95\% 
            \\
            \midrule[1.5pt]
            \multirow{11}{*}{\rotatebox{90}{\bf{Predictor-Based}}}
            &{CNN} & \multirow{7}{*}{\XSolidBrush} & 0.366 & 0.372 & 0.545 && 0.197 & 0.197 & 0.244 && 0.238 & 0.256 & 0.354 && 0.258 & 0.260 & 0.387 \\
            &{ResNet} & & 0.409  & 0.450 & 0.605 && 0.230  & 0.234  & 0.280 && 0.282 & 0.308 & 0.405 && 0.277  & 0.280 & 0.304 \\
            &{LSTM} & & 0.247  & 0.270  & 0.425 && 0.194  & 0.195  & 0.225 && 0.223 &  0.245 & 0.321 && 0.263  & 0.269 & 0.283 \\
            &{Transformer} &  & 0.167 & 0.175 & 0.238 && 0.267  & 0.262 & 0.264 && 0.184 & 0.195 & 0.211 && 0.378  & 0.388 & 0.405 \\
            &{GCN} &  & 0.245 & 0.246 & 0.320 && 0.251 & 0.248  & 0.252 && 0.180 & 0.187 & 0.195 && 0.318 & 0.320 & 0.329\\
            &{GearNet-P} & &\bf{0.700} & \bf{0.769} & \bf{0.854} && 0.348 & 0.359 & 0.406 && 0.482 & 0.525 & 0.613 && 0.407 & 0.418 & 0.458\\
            &{CDConv-P} &  & {0.634} & {0.702} & {0.820} && \bf{0.381} & \bf{0.401} & \bf{0.453} && \bf{0.533} & \bf{0.577} & \bf{0.654} && {\bf{0.428}} & {\bf{0.440}} & {\bf{0.479}} \\
            \cmidrule{2-18}
            &{DeepFRI} & \multirow{4}{*}{\Checkmark} & 0.470 & 0.545 & 0.631 && 0.361 & 0.371 & 0.399 && 0.374 & 0.409 & 0.465 && 0.440 & 0.444 & 0.460 \\
            &{ProtBERT-BFD} &  & 0.691 & 0.752 & 0.838 && 0.308 & 0.321 & 0.361 && 0.497 & 0.541 & 0.613 && 0.287 & 0.293 & 0.308 \\
            &{ESM-2-650M-P} &  & {0.763} & {0.816} & {0.877} && 0.423 & 0.438 & {0.484} && 0.563 & 0.604 & 0.661 && \textcolor{blue}{\bf 0.497} & \textcolor{blue}{\bf 0.509} & \textcolor{blue}{\bf 0.535}\\
            &{PromptProtein} & & \textcolor{blue}{\bf{0.765}} & \textcolor{blue}{\bf{0.823}} & \textcolor{red}{\bf{0.888}} && \textcolor{red}{\bf{0.439}} & \textcolor{red}{\bf{0.453}} & \textcolor{red}{\bf{0.495}} && \textcolor{red}{\bf{0.577}} & \textcolor{red}{\bf{0.600}} & \textcolor{red}{\bf{0.677}} && \textcolor{red}{\bf{0.532}} & \textcolor{red}{\bf{0.533}} & \textcolor{red}{\bf{0.551}} \\
            \midrule[1.5pt]
            \multirow{10}{*}{\rotatebox{90}{\bf{Retriever-Based}}}
            & {MMseqs} & \multirow{9}{*}{\XSolidBrush} & \textcolor{red}{\bf{0.781}} & \textcolor{red}{\bf{0.833}} & \textcolor{blue}{\bf{0.887}} && 0.323 & 0.359 & 0.444 && 0.502 & 0.557 & 0.647 && 0.237 & 0.255 & 0.332 \\
            & {BLAST} & & 0.740 & 0.806 & 0.872 && 0.344 & 0.373 & 0.448 && 0.505 & 0.557 & 0.640 && 0.275 & 0.284 & 0.347 \\
            & {PSI-BLAST} & & 0.780 & 0.822 & 0.885 && 0.323 & 0.352 & 0.433 && 0.509 & 0.558 & 0.641 && 0.242 & 0.253 & 0.324 \\
            & {TMAlign} & & 0.674 & 0.744 & 0.817 && 0.403 & 0.426 & 0.480 && 0.487 & 0.533 & 0.597 && \bf{0.410} & \bf{0.424} & \bf{0.456}\\
            & {Foldseek} & & {0.781} & {0.834} & {0.885} && 0.328 &0.359 & 0.435 && 0.525 &0.573 &0.651 && 0.245 & 0.254 & 0.312\\
            & {Progres} & & 0.535 & 0.634 & 0.727 && 0.353 & 0.379 & 0.448 && 0.428 & 0.480 & 0.573 && 0.374 & 0.390 & 0.438\\
            & {GearNet-R} & & 0.671 & 0.744 & 0.822 && 0.391 & 0.419 & 0.482 && 0.497 & 0.548 & 0.626 && 0.377 & 0.387 & 0.434\\
            & {CDConv-R} & & 0.719 & 0.784 & 0.843 && \textcolor{blue}{\bf{0.409}} & \textcolor{blue}{\bf{0.434}} &	\textcolor{blue}{\bf{0.494}}&& \bf{0.536} & \bf{0.584} & \bf{0.661} && 0.387 & 0.397 &	0.438\\
            \cmidrule{2-18}
            & {ESM-2-650M-R} & \multirow{2}{*}{\Checkmark} & 0.585 & 0.656 & 0.753 && \bf{0.398} & \bf{0.415} & \bf{0.477} && 0.462 & 0.510 &	0.607 && {\bf{0.427}} & {\bf{0.436}} & {\bf{0.472}}\\
            & {TM-Vec} &  & \bf{0.676} & \bf{0.745} & \bf{0.817} && 0.377 & 0.399 & 0.461 && \textcolor{blue}{\bf{0.552}} & \textcolor{blue}{\bf{0.593}} & \textcolor{blue}{\bf{0.663}} && 0.328 & 0.328 & 0.369\\
            \bottomrule[2pt]
        \end{tabular}
    \end{adjustbox}
    \begin{tablenotes}
        \item[*] \footnotesize \textcolor{red}{\bf{Red}}: the best results among all; \textcolor{blue}{\bf{blue}}: the second best results among all; \textbf{bold}: the best results within blocks.
    \end{tablenotes}
    \end{threeparttable}
\vspace{-0.2cm}
\end{table*}

\subsection{Detailed Benchmarking Results}
\label{app:sec:benchmark}
The detailed benchmark results on EC and GO are shown in Tab.~\ref{tab:all_result}.

\subsection{Applying Retriever to Real-World Function Annotation}
\label{app:exp:clean}

In addition to the benchmark results presented in Sec.~\ref{sec:exp:benchmark}, we now extend to studies that explore EC number prediction under more real-world and challenging settings~\citep{yu2023enzyme,sanderson2021proteinfer}. 
Specifically, CLEAN~\citep{yu2023enzyme} introduces a contrastive supervised learning approach that aligns protein representations with analogous enzyme commission numbers, an approach which has been substantiated through empirical validation.
In this work, we deploy our proposed retrievers on their test sets without function-specific training on their respective training sets. This methodological choice is made to demonstrate the effectiveness of our retrieval-based approach in realistic settings.

\textbf{Setup.}
We closely follow CLEAN~\citep{yu2023enzyme} settings for evaluation.
Baselines are trained on or retrieved against the collected Swiss-Prot dataset in~\citet{yu2023enzyme} with 227,363 protein sequences.
Two independent test sets are used for a fair and rigorous benchmark study.
The first, an enzyme sequence dataset, includes 392 sequences that span 177 different EC numbers. These sequences were released post-April 2022, subsequent to the proteins in our training set, reflecting a real-world scenario where the Swiss-Prot database serves as the labeled knowledge base, and the functions of the query sequences remain unidentified.
The second test set, known as Price-149, consists of experimentally validated findings detailed by~\citet{price2018mutant}. This dataset, initially curated by~\citet{sanderson2021proteinfer} as a benchmark for challenge, features sequences that were previously mislabeled or inconsistently annotated in automated systems.

\textbf{Methods.}
We select four EC number annotation tools as baselines: CLEAN~\citep{yu2023enzyme}, ProteInfer~\citep{sanderson2021proteinfer}, ECPred~\citep{dalkiran2018ecpred}, DeepEC~\citep{ryu2019deep}, the results of which are directly taken from the CLEAN paper~\citep{yu2023enzyme}.
For comparison, we test the performance of traditional and neural retrievers considered in Sec.~\ref{sec:exp:benchmark}.
Due to the large size of Swiss-Prot training set, we do not consider predictor-based methods and the ProtIR framework that requires training.
This decision allows for a focused comparison on the effectiveness of retrieval-based approaches.
We also ignore the TM-Align method, as it takes over 24 hours to finish the retrieval on Swiss-Prot.

It is important to note that structure-based retrievers, such as GearNet-R and CDConv-R, require protein structures for input, which are not experimentally available for most proteins in Swiss-Prot. However, with the advent of the AlphaFold Database~\citep{varadi2022alphafold}, accurate structure predictions for the Swiss-Prot proteins made by AlphaFold2 are now accessible. For the purposes of our model, we search the available structures directly from the AlphaFold Database, successfully retrieving structures for 224,515 out of 227,363 proteins in the training (retrieved) set. A similar approach was adopted for the NEW-392 and Price-149 test sets, from which all structures were obtained.
For those structures not in AlphaFold Database, we run ColabFold~\citep{mirdita2022colabfold} to get their predicted structures.

\begin{table}[t]
    \centering
    \caption{Results on NEW-392 and Price-149 test sets of EC annotation tools with (a) function-specific training , (b) traditional retrievers and (c) neural retrievers. The results in the (a) block are taken from~\citet{yu2023enzyme}. Retriever-based methods, even without function-specific training, consistently outperform EC annotation tools.}
    \vspace{-0.8em}
    \label{tab:clean}
    \begin{adjustbox}{max width=0.6\linewidth}
        \begin{tabular}{lcccccccc}
            \toprule[2pt]
            &\multirow{2}{*}{{\bf{Method}}} &
            \multicolumn{3}{c}{\bf{NEW-392}}&&
            \multicolumn{3}{c}{\bf{Price-149}} 
            \\
            \cmidrule[1pt]{3-5}
            \cmidrule[1pt]{7-9}
            & & Precision & Recall & F1 &&
            Precision & Recall & F1
            \\
            \midrule[1.5pt]
            \multirow{4}{*}{(a)} & ECPred & 0.117	& 0.095 & 0.100 && 0.019 & 0.019 & 0.019\\
            & DeepEC & 0.297 & 0.216 & 0.229 &&0.118 & 0.072 & 0.084 \\
            & ProteInfer & 0.408 & 0.284 & 0.308 && 0.243 & 0.138 & 0.166 \\
            & CLEAN & 0.596 & 0.481 & 0.498 &&{0.584} & 0.467 & 0.494 \\
            \midrule[1pt]
            \multirow{4}{*}{(b)} & BLAST & 0.593 & 0.648 & 0.585 && 0.538 & 0.480 & 0.477 \\
            & PSI-BLAST & \bf{0.675} & 0.682 & 0.625 && 0.577 & \bf{0.559} & \bf{0.536} \\
            & Foldseek & {0.666} & 0.680 & 0.612 && 0.537 & 0.500 & 0.484 \\
            & MMseqs & {0.624} & \bf{0.700} & {0.632} && {0.583} & {0.546} & {0.534} \\
            \midrule
            \multirow{5}{*}{(c)} & Progres & 0.525 & 0.557 & 0.513 && 0.519 & 0.434 & 0.430 \\
            & TM-Vec &{0.634} & \bf{0.700} & \bf{0.639} && \bf{0.591} & 0.533 & 0.524 \\
            & GearNet-R & 0.636 & 0.668 & 0.597 && 0.531 & 0.487 & 0.469 \\
            & CDConv-R & 0.588 & 0.618 & 0.577 && 0.546 & 0.487 & 0.463 \\
            & ESM-2-650M-R & 0.594 & 0.602 & 0.537	&& 0.533 & 0.474 & 0.478 \\
            \bottomrule[2pt]
        \end{tabular}
    \end{adjustbox}
\end{table}

\textbf{Results.}
The results are shown in Tab.~\ref{tab:clean}. 
First, it is evident that all considered retrievers surpass the performance of CLEAN on the NEW-392 test set in F1 score, despite not undergoing any function-specific training on the training set—a process that CLEAN underwent. This underscores the potency of retriever-based approaches.
Second, despite the lack of experimentally determined structures, structure retrievers, \emph{e.g.}, Foldseek and GearNet-R, demonstrate high performance with AlphaFold2-predicted structures on both considered test sets. Here, GearNet-R exhibits superior performance over the supervised retriever CLEAN and the PLM-based retriever ESM-2-650M.
This proves that structure-based retrievers work well with accurately predicted structures, which are easier to obtain than experimental structures.
Aligning with our observation in Sec.~\ref{sec:exp:benchmark}, sequence-based methods excel on EC number annotation tasks than structure-based methods, which highlights the importance of evolutionary information.
To conclude, retriever-based methods continue to demonstrate their potential in practical scenarios, emphasizing the critical role of modeling similarities between proteins.

\subsection{Comparison between \method and ensemble baselines}

To demonstrate the efficacy of the \method framework, we conducted a comparison involving the \method-augmented GearNet and CDConv predictors against a basic ensemble baseline. This ensemble approach involves averaging the predictions made by the predictor and its corresponding retriever, with the results presented in Tab.~\ref{tab:ensemble}.

The results in the table reveal that while ensembling serves as a robust baseline for most tasks, our method is able to consistently enhance this baseline, achieving an improvement in the range of approximately 2\% to 4\% in terms of F\textsubscript{max}. This improvement highlights the added value and effectiveness of the \method framework in enhancing prediction accuracy across various tasks.

\begin{table*}[h]
    \centering
    \caption{F\textsubscript{max} on EC and GO prediction with \method and ensemble baselines, where the former performs better.
    }
    \label{tab:ensemble}
    \begin{threeparttable}
    \begin{adjustbox}{max width=\linewidth}
        \begin{tabular}{l|ccccccccccccccccc}
            \toprule[2pt]
            \multirow{2}{*}{{\bf{Model}}} & \multirow{2}{*}{{\bf{Method}}} &
            \multicolumn{3}{c}{\bf{EC}}&&
            \multicolumn{3}{c}{\bf{GO-BP}} && 
            \multicolumn{3}{c}{\bf{GO-MF}} && 
            \multicolumn{3}{c}{\bf{GO-CC}}
            \\
            \cmidrule{3-5}
            \cmidrule{7-9}
            \cmidrule{11-13}
            \cmidrule{15-17}
            & & 30\% & 50\% & 95\% &&
            30\% & 50\% & 95\% && 
            30\% & 50\% & 95\% && 
            30\% & 50\% & 95\% 
            \\
            \midrule[1.5pt]
            \multirow{4}{*}{\bf{GearNet}} & {Predictor} & 0.700 & 0.769 & 0.854 && 0.348 & 0.359 & 0.406 && 0.482 & 0.525 & 0.613 && 0.407 & 0.418 & 0.458 \\
            & {Retriever} & 0.671 & 0.744 & 0.822 && 0.391 & 0.419 & 0.482 && 0.497 & 0.548 & 0.626 && 0.377 & 0.387 & 0.434\\
            & {Ensemble} &0.720 & 0.797 & 0.861 && 0.394 & 0.421 & 0.486 && 0.512 & 0.551 & 0.630 && 0.423 & 0.437 & 0.464 \\
            \cmidrule{2-17}
            & \bf{\method} & \bf{0.743} & \bf{0.810} & \bf{0.881} && \bf{0.409} & \bf{0.431} &	\bf{0.488}&& \bf{0.518} & \bf{0.564} &	\bf{0.650}&& \bf{0.439} & \bf{0.452} &	\bf{0.501} \\
            \midrule[1.5pt]
            \multirow{4}{*}{\bf{CDConv}} & {Predictor} & 0.634 & 0.702 & 0.820 && 0.381 & 0.401 & 0.453 && 0.533 & 0.577 & 0.654 && 0.428 & 0.440 & 0.479 \\
            & {Retriever} & 0.719 & 0.784 & 0.843 && {0.409} & {0.434} &	{0.494}&& {0.536} & {0.584} & {0.661} && 0.387 & 0.397 &	0.438\\
            & {Ensemble} & 0.724 & 0.802 & 0.864 && 0.414 & 0.438 & 0.495 && 0.555 & 0.596 & 0.665 && 0.431 & 0.443 & 0.478 \\
            \cmidrule{2-17}
            & \bf{\method} & \bf{0.769} & \bf{0.820} & \bf{0.885} && \bf{0.434} & \bf{0.453} &	\bf{0.503}&& \bf{0.567} & \bf{0.608} &	\bf{0.678}&& \bf{0.447} & \bf{0.460} &	\bf{0.499} \\
            \midrule[1.5pt]
            \multicolumn{2}{c}{{PromptProtein}} & {0.765} & {0.823} & {0.888} && {0.439} & {0.453} & {0.495} && {0.577} & {0.600} & {0.677} && {0.532} & {0.533} & {0.551} \\
            \bottomrule[2pt]
        \end{tabular}
    \end{adjustbox}
    \end{threeparttable}
\end{table*}

\subsection{Hyperparameter Configuration}

\begin{figure*}[h]
    \centering
    \includegraphics[width=0.8\linewidth]{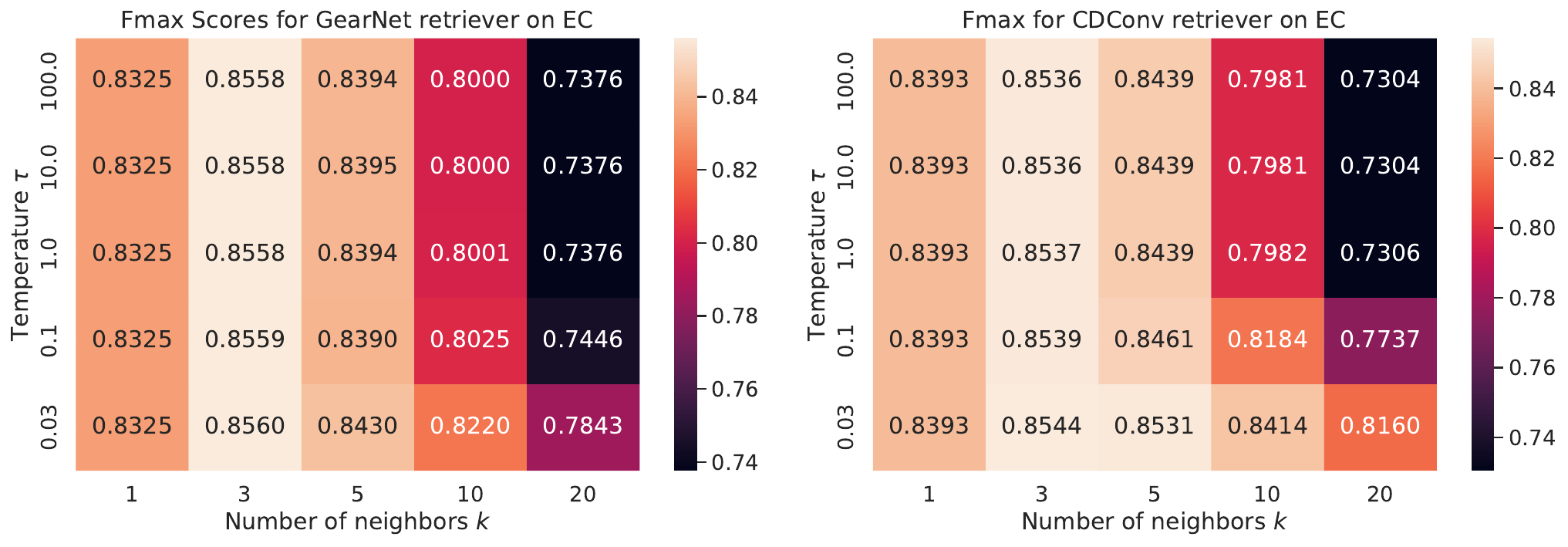}
    \caption{Change of F\textsubscript{max} on EC with respect to $k$ and $\tau$ in Eqs. (\ref{eq:retriver})(\ref{eq:kernel}) for two retrievers.}
    \label{fig:ktemp}
\end{figure*}

\textbf{Hyperparameter analysis for retriever-based methods.}
To investigate the impact of the number of retrieved neighbors ($k$) and the temperature parameter ($\tau$) on the performance of function annotation in retriever-based methods, we plot a heatmap for this hyperparameter analysis, as shown in Fig.~\ref{fig:ktemp}. We observe that a temperature of $\tau=0.03$ yields the most effective results for scaling the cosine similarity between protein representations. This optimal setting can be attributed to the nature of cosine similarity, which ranges between $[-1, 1]$; without amplification by the temperature, there is minimal variation in the weights assigned to different proteins.

Furthermore, we note that at lower values of $k$, the effect of the temperature parameter is relatively minor, primarily because most of the retrieved proteins tend to have the same EC number. However, as $k$ increases, leading to a wider variety of retrieved EC numbers, the temperature becomes more influential. In such scenarios, it serves to emphasize proteins that are more similar to the query protein, thereby refining the function annotation process. This understanding highlights the importance of carefully selecting the values of $k$ and $\tau$ to optimize the performance of retriever-based methods.

\textbf{Hyperparameter tuning for \method framework.}
The tuning process for the \method framework is divided into two main stages: the pre-training stage and the refinement stage.

In the pre-training stage, for both predictors and retrievers, we adhere to the optimal hyperparameters established in prior research~\citep{zhang2022protein}. This includes settings for the learning rate, batch size, and the number of epochs. The model that achieves the best performance on the validation set is then selected to proceed to the refinement stage.

During the refinement stage, the predictor and retriever are iteratively refined. In each iteration, it is crucial to balance the models' convergence with the goal of fitting pseudo-labels, while also being mindful of potential overfitting. To maintain this balance, we closely monitor performance metrics on the validation set and halt training when no further improvements are observed. Notably, test set performance is not considered during training to ensure a fair comparison.

Based on our experience, training for approximately 30 epochs during both the E-step and M-step is typically sufficient for the convergence of both the predictor and retriever. Moreover, the validation performance often stabilizes after around five rounds. The final step involves selecting the model with the best performance on the validation set and subsequently evaluating it on the test set.

\end{document}